\documentstyle[aps]{revtex}
\begin{document}
\draft
\title{Gravitational-wave dynamics and black-hole dynamics:\\
second quasi-spherical approximation}
\author{Sean A. Hayward}
\address{Asia Pacific Center for Theoretical Physics,\\
The Korea Foundation for Advanced Study Building 7th Floor,
Yoksam-dong 678-39, Kangnam-gu, Seoul 135-081, Korea\\
and\\
Department of Physics, Konkuk University,
93-1 Mojin-dong, Kwangjin-gu, Seoul 143-701, Korea\\
{\tt hayward@mail.apctp.org}}
\date{2nd February 2001}
\maketitle

\begin{abstract}
Gravitational radiation with roughly spherical wavefronts,
produced by roughly spherical black holes or other astrophysical objects,
is described by an approximation scheme.
The first quasi-spherical approximation, 
describing radiation propagation on a background,
is generalized to include additional non-linear effects, 
due to the radiation itself.
The gravitational radiation is locally defined and admits an energy tensor,
satisfying all standard local energy conditions
and entering the truncated Einstein equations as an effective energy tensor.
This second quasi-spherical approximation thereby includes 
gravitational radiation reaction,
such as the back-reaction on the black hole.
With respect to a canonical flow of time,
the combined energy-momentum of the matter and gravitational radiation
is covariantly conserved.
The corresponding Noether charge is a local gravitational mass-energy.
Energy conservation is formulated as a local first law relating 
the gradient of the gravitational mass to work and energy-supply terms,
including the energy flux of the gravitational radiation.
Zeroth, first and second laws of black-hole dynamics are given,
involving a dynamic surface gravity.
Local gravitational-wave dynamics is described by a non-linear wave equation.
In terms of a complex gravitational-radiation potential,
the energy tensor has a scalar-field form 
and the wave equation is an Ernst equation, 
holding independently at each spherical angle.
The strain to be measured by a distant detector is simply defined.
\end{abstract}

\section{Introduction}

Gravitational waves and black holes 
are among the most popular and intensively investigated topics in physics 
at the turn of the millenium.
Both are predictions of Einstein's General Relativity\cite{E1}
which involve essentially relativistic gravitational effects.
Gravitational waves were originally introduced by Einstein himself\cite{E2}
and the first black-hole solution 
was also almost immediately discovered\cite{S},
though the term black hole was coined much later\cite{W1}.
In the remainder of the twentieth century,
astrophysical evidence eventually accumulated to the point where 
it is nowadays believed that black holes are not only common in the universe,
but astrophysically dominant energy sources and appreciable mass concentrations,
being the final remnant of any sufficiently massive star,
with supermassive black holes powering active galactic nuclei
and lurking at the heart of most other galaxies, e.g.\cite{HK}.
Gravitational waves are expected to make a similar transition from theory 
to observation with the operation of 
several new gravitational-wave detectors, e.g.\cite{GLPPS}.
Observation of gravitational waves from black holes 
would provide the first direct evidence for the existence of the latter,
rather than the impressive but indirect evidence 
of their effect on other astrophysical bodies or surrounding matter.
The combination of these two intertwined topics,
gravitational waves and black holes,
poses an exceptional challenge for theorists.

For instance, an expected source of detectable gravitational waves is 
the inspiral and coalescence of a binary black-hole system.
Although the earliest and latest stages are understood 
in terms of post-Newtonian\cite{Bl} 
and close-limit approximations\cite{P} respectively,
the coalescence is generally thought to be tractable only by 
full numerical simulations\cite{BBH1,BBH2}.
Textbook theory simply does not suffice to understand the process 
in physical terms.
Stationary black holes are well understood\cite{BCH,Wa,C},
but the black holes of interest are highly dynamical.
Asymptotic or weak gravitational waves are well understood,
respectively by Bondi-Penrose theory\cite{B,BBM,S1,S2,P1,P2,PR} 
and Wheeler-Isaacson high-frequency 
(or merely linearized) theory\cite{W2,PW,BH,I,MTW,T},
but strong gravitational waves produced by a distorted, 
rapidly evolving black hole are not understood at all.
Indeed, it is sometimes argued that there is nothing to physically understand,
merely complicated equations to numerically integrate,
as gravitational waves cannot be localized\cite{MTW}.
The main purpose of this article is to address this lack of relevant theory
by providing an astrophysically realistic approximation scheme
in which both gravitational radiation and black holes are locally defined, 
along with their physical attributes,
with each dynamically influencing the other.
The approximation can be simply stated:
it holds where the gravitational wavefronts 
and black hole (or other astrophysical object) are roughly spherical.
This does not contradict 
the classically quadrupolar nature of gravitational waves,
as the wave amplitude at different angles may be arbitrarily aspherical.
The wavefronts, not the waves themselves, should be roughly spherical.

With such a general premise of rough sphericity, 
this quasi-spherical approximation scheme 
intuitively promises a wide range of validity,
in particular including a just-coalesced black hole.
It may also prove applicable to neutron stars or supernovas,
or indeed any astrophysical situation which has rough spherical symmetry.
Further, 
for any other process which can be enclosed by roughly spherical surfaces,
it provides a mid-zone and, assuming isolation, far-zone approximation.
Mathematically this was originally achieved\cite{qs} 
by linearizing certain fields which would vanish in exact spherical symmetry,
having made a decomposition of space-time 
adapted to the roughly spherical surfaces,
henceforth called transverse surfaces.
This will henceforth be called the {\em first quasi-spherical approximation}.
Since there is no assumption of closeness to stationarity,
the approximation holds for arbitrarily fast dynamical processes.
The approximation has also been tested against 
the main expected source of asphericity, angular momentum,
by applying it to Kerr black holes\cite{SH}:
the error in the strain waveform is much lower than 
expected signals from binary black-hole coalescence.

The approximation allows a local definition of gravitational radiation 
essentially because the decomposed fields naturally divide 
into quasi-spherical variables and wave variables, 
the latter satisfying a wave equation and yielding the Bondi news, 
or equivalently, the observable strain.
Specifically, the gravitational radiation is encoded in 
the shear tensors of the outgoing and ingoing wavefronts.
Then by retaining non-linear terms in these (previously linearized) fields,
one would expect to obtain an approximation 
which is more accurate for the gravitational-radiation sector.
This {\em second quasi-spherical approximation} is also presented here.
Both first and second approximations share the remarkable feature that, 
to compute the observable waveforms,
no transverse derivatives need be considered.
The truncated equations form an effectively two-dimensional system,
already written in characteristic form 
by virtue of the first-order dual-null formulation, 
to be integrated independently at each angle of the sphere.
Numerically this is much easier to implement 
and computationally inexpensive when compared to the full Einstein system.
Numerical codes exist for both first and second approximations\cite{SH}.
Moreover, the wave equation can be written as an Ernst equation
in terms of a gravitational-radiation potential.
Thus the {\em gravitational-wave dynamics} is amenable to analytical methods.

The first approximation describes 
gravitational-wave propagation on a background,
though the background is neither necessarily spherical nor fixed in advance;
this merely means that 
the quasi-spherical equations decouple from the gravitational-wave equation.
There is no such decoupling in the second approximation,
and therefore no background which is independent of the waves.
Back-reaction of the waves on the geometry is thereby included.
Specifically, 
one may define a {\em gravitational-radiation energy tensor}\cite{gwe} 
which acts just like a matter energy tensor in the truncated Einstein equations.
Thus there is a fully relativistic inclusion of 
{\em gravitational radiation reaction} for dynamic black holes.
For instance, if a black hole emits gravitational radiation, 
it backscatters to produce ingoing radiation 
which is absorbed by the black hole,
thereby increasing its mass and area.
This last property follows from a local second law\cite{bhd,bhs},
part of a general theory of {\em black-hole dynamics}\cite{mg9},
where a black hole is defined by a type of trapping horizon.
This local theory of dynamic black holes,
not to be confused with textbook black-hole statics and asymptotics,
is extended here by deriving quasi-spherical generalizations of 
the spherically symmetric first\cite{1st} 
and zeroth\cite{mg9} laws of black-hole dynamics,
involving local definitions of {\em mass} and {\em surface gravity}.

The article is organized as follows.
\S II reviews the formalism of dual-null dynamics,
which provides a geometrical description of the wavefronts
and the corresponding decomposition of Einstein gravity.
\S III describes both first and second quasi-spherical approximations
and the resulting truncated field equations.
\S IV locally defines the gravitational radiation 
and its energy tensor $\Theta$.
A truncated Einstein tensor illustrates the role of $\Theta$
as an effective matter energy tensor in the second approximation.
\S V locally defines dynamic black holes,
along with mass $m$ and surface gravity $\kappa$.
\S VI derives an energy conservation law 
for the combined energy-momentum of the matter and gravitational radiation.
This is also formulated as a first law relating the gradient of $m$
to work and energy-supply terms,
including the energy flux of the gravitational radiation.
\S VII gives zeroth, first and second laws of black-hole dynamics
and various inequalities.
\S VIII introduces the complex gravitational-radiation potential,
in terms of which $\Theta$ takes a scalar-field form 
and the gravitational-wave equation takes an Ernst form.
\S IX defines conformally rescaled fields more suited to asymptotics,
including a localized Bondi flux and a conformal strain tensor.
\S X concludes.

\section{Dual-null dynamics}

The wavefronts of outgoing and ingoing gravitational radiation 
form two families of null hypersurfaces,
intersecting in the two-parameter family of transverse spatial surfaces.
This geometry is described by the formalism of dual-null dynamics\cite{dn,dne},
summarized in this section. 
Denoting the space-time metric by $g$
and labelling the null hypersurfaces by $x^\pm$, 
the normal 1-forms $n^\pm=-dx^\pm$ therefore satisfy
\begin{equation}
g^{-1}(n^\pm,n^\pm)=0.
\end{equation}
The relative normalization of the null normals 
may be encoded in a function $f$ defined by
\begin{equation}
e^f=-g^{-1}(n^+,n^-).
\end{equation}
Then the induced metric on the transverse surfaces,
the spatial surfaces of intersection, is found to be
\begin{equation}
h=g+2e^{-f}n^+\otimes n^-
\end{equation}
where $\otimes$ denotes the symmetric tensor product.
The dynamics is generated by two commuting evolution vectors $u_\pm$:
\begin{equation}
[u_+,u_-]=0
\end{equation}
where the brackets denote the Lie bracket or commutator.
Thus there is an integrable evolution space spanned by $(u_+,u_-)$.
The evolution derivatives, to be discretized in a numerical code, 
are the projected Lie derivatives
\begin{equation}
\Delta_\pm=\bot L_{u_{\pm}}
\end{equation}
where $\bot$ indicates projection by $h$ and $L$ denotes the Lie derivative.
There are two shift vectors 
\begin{equation}
s_\pm=\bot u_\pm.
\end{equation}
In a coordinate basis $(u_+,u_-,e_1,e_2)$ such that
$u_\pm=\partial/\partial x^\pm$, 
where $e_a=\partial/\partial x^a$ is a basis for the transverse surfaces,
the metric takes the form
\begin{equation}
g=h_{ab}(dx^a+s_+^adx^++s_-^adx^-)\otimes(dx^b+s_+^bdx^++s_-^bdx^-)
-2e^{-f}dx^+\otimes dx^-.
\end{equation}
Then $(h,f,s_\pm)$ are configuration fields
and the independent momentum fields are found to be linear combinations of
\begin{eqnarray}
\theta_\pm&=&\bar{*}L_\pm\bar{*}1\\
\sigma_\pm&=&\bot L_\pm h-\theta_\pm h\\
\nu_\pm&=&L_\pm f\\
\omega&=&\textstyle{1\over2}e^fh([l_-,l_+])
\end{eqnarray}
where $\bar{*}$ is the Hodge operator of $h$ 
and $L_\pm$ is shorthand for the Lie derivative along the null normal vectors 
\begin{equation}
l_\pm=u_\pm-s_\pm=e^{-f}g^{-1}(n^\mp)
\end{equation}
which will be assumed to be future-null.
Then the functions $\theta_\pm$ are the expansions,
the traceless bilinear forms $\sigma_\pm$ are the shears,
the 1-form $\omega$ is the twist,
measuring the lack of integrability of the normal space,
and the functions $\nu_\pm$ are the inaffinities, 
measuring the failure of the null normals to be affine. 
The fields $(\theta_\pm,\sigma_\pm,\nu_\pm,\omega)$ 
encode the extrinsic curvature of the dual-null foliation.
These extrinsic fields are unique up to duality $\pm\mapsto\mp$
and diffeomorphisms which relabel the null hypersurfaces, i.e.\
$dx^\pm\mapsto e^{\lambda_\pm}dx^\pm$
for functions $\lambda_\pm(x^\pm)$.

The dual-null Hamilton equations and integrability conditions 
for vacuum Einstein gravity have been given previously
in a different notation\cite{dne}.
They are linear combinations of the vacuum Einstein equation
and a first integral of the contracted Bianchi identity. 
It is straightforward to add matter:
denoting the projections of the energy tensor $T$ by
$T_{\pm\pm}=T(l_\pm,l_\pm)$,
$T_{+-}=T(l_+,l_-)$ and
$T_\pm=\bot T(l_\pm)$,
the resulting field equations are
\begin{eqnarray}
L_\pm\theta_\pm&=&-\nu_\pm\theta_\pm-\textstyle{1\over2}\theta_\pm^2
-\textstyle{1\over4}|\sigma_\pm|^2
-8\pi T_{\pm\pm}
\\
L_\pm\theta_\mp&=&-\theta_+\theta_-
-e^{-f}\left(
\textstyle{1\over2}R-|\textstyle{1\over2}D f\pm\omega|^2
+D\cdot(\textstyle{1\over2}D f\pm\omega)^\sharp\right)
+8\pi T_{+-}
\\
L_\pm\nu_\mp&=&\textstyle{1\over4}\sigma_+:\sigma_-^\sharp
-\textstyle{1\over2}\theta_+\theta_-
-e^{-f}\left(
\textstyle{1\over2}R-3|\omega|^2+\textstyle{1\over4}|D f|^2
\pm\omega^\sharp\cdot D f\right)
+8\pi\left(T_{+-}+\textstyle{1\over2}e^{-f}h^\sharp:T\right)
\\
\bot L_\pm\omega&=&-\theta_\pm\omega
\pm\textstyle{1\over2}(D^\sharp\cdot\sigma_\pm
-D\theta_\pm+D\nu_\pm-\theta_\pm Df)
\mp8\pi T_\pm
\\
\bot L_\pm\sigma_\mp&=&\sigma_+\cdot h^\sharp\cdot\sigma_-
\pm\textstyle{1\over2}(\theta_+\sigma_--\theta_-\sigma_+)
+2e^{-f}\left((\textstyle{1\over2}Df\pm\omega)\otimes
(\textstyle{1\over2}Df\pm\omega)
-D\otimes(\textstyle{1\over2}Df\pm\omega)\right)
\nonumber\\
&&-e^{-f}\left(|\textstyle{1\over2}Df\pm\omega|^2
-D\cdot(\textstyle{1\over2}Df\pm\omega)^\sharp\right)h
+8\pi e^{-f}\left(\bot T-\textstyle{1\over2}(h^\sharp:T)h\right)
\\
\bot\left(L_+s_--L_-s_+\right)&=&2e^{-f}h^\sharp\cdot\omega+[s_-,s_+]
\\
\bot L_\pm h&=&\theta_\pm h+\sigma_\pm
\\
L_\pm f&=&\nu_\pm
\end{eqnarray}
where $D$ is the covariant derivative and $R$ the Ricci scalar of $h$,
a dot denotes symmetric contraction,
a colon denotes double symmetric contraction,
a sharp ($\sharp$) denotes the contravariant dual 
with respect to $h^{-1}=h^\sharp$ (index raising),
a flat ($\flat$) will denote the covariant dual 
with respect to $h$ (index lowering),
$|\omega|^2=\omega\cdot\omega^\sharp$
and $|\sigma|^2=\sigma:\sigma^\sharp$.
Units are such that Newton's gravitational constant is unity.
This is the Einstein system in first-order dual-null form;
the equations will simply be called the field equations.
The second-order version obtained by eliminating the momentum fields 
and symmetrizing in $L_{(+}L_{-)}$ are the Einstein equations themselves.

\section{Quasi-spherical approximations}

The dual-null fields and operators fall into two classes,
those which vanish in spherical symmetry, $(\sigma_\pm,\omega,s_\pm,D)$, 
and those which generally do not,
$(\theta_\pm,\nu_\pm,h,f,\Delta_\pm)$\cite{1st,sph}.
The {\em first quasi-spherical approximation}\cite{qs} 
therefore consisted of linearizing in $(\sigma_\pm,\omega,s_\pm,D)$. 
It was then noticed that the gravitational radiation is encoded in 
the shear tensors $\sigma_\pm$\cite{gwe},
as will be explained in more detail in the following.
This suggests that retaining non-linear terms in $\sigma_\pm$
would give a more accurate approximation 
for the gravitational-radiation sector of the theory.
Thus the {\em second quasi-spherical approximation} 
consists of linearizing in $(\omega,s_\pm,D)$ only. 
For a given matter model, 
one would also have to decide which decomposed matter fields to linearize.
In this article, no specific matter model will be fixed,
but the matter energy tensor $T$ will be retained for generality 
and assumed to be consistently truncated.

More formally, 
one may introduce two expansion parameters into the full field equations,
$\epsilon_1$ preceding $\sigma_\pm$ 
and $\epsilon_0$ preceding $(\omega,s_\pm,D)$.
Then both approximations ignore terms $o(\epsilon_0)$,
whereas the first approximation ignores terms $o(\epsilon_1)$.
Then $\epsilon_1$ measures the strength of the gravitational radiation,
whereas $\epsilon_0$ measures other asphericities,
due to angular momentum or other transverse effects.
The terminology ``first and second order'' has been carefully avoided because 
the true second-order quasi-spherical approximation 
would be full Einstein gravity, 
since the only non-linear terms in the full field equations 
are second-order in the dynamical fields and operators.
However, one may say that 
the second approximation is second-order in the gravitational radiation.

It is useful to decompose the transverse metric $h$ into 
a conformal factor $r$ and a {\em transverse conformal metric} $k$ by
\begin{equation}
h=r^2k
\end{equation}
such that 
\begin{equation}
\Delta_\pm\hat{*}1=0
\end{equation}
where $\hat{*}$ is the Hodge operator of $k$, satisfying $\bar{*}1=\hat{*}r^2$.
The Ricci scalar of $h$ is found to be
\begin{equation}
R=2r^{-2}(1-D^2\ln r)
\end{equation}
by using the coordinate freedom on a given surface 
to fix $k$ as the metric of a unit sphere.
One may take quasi-spherical coordinates $x^a=(\vartheta,\varphi)$
on the transverse surfaces to obtain the standard area form of a unit sphere:
\begin{equation}
\hat{*}1=\sin\vartheta\,d\vartheta\wedge d\varphi
\end{equation}
where $\wedge$ denotes the exterior product of forms.
Then $r$ is the {\em quasi-spherical radius}.
Approximations for rough cylindrical or plane symmetry 
could similarly be produced.
A useful truncation identity, 
holding in both first and second approximations, is 
\begin{equation}
\Delta_\pm=\bot L_\pm.
\end{equation}
This will be used throughout the article without further reference,
with $\Delta_\pm$ rather than $L_\pm$ appearing explicitly.

In the first approximation, 
the truncated field equations decouple into a three-level hierarchy.
The first-level equations 
\begin{eqnarray}
\label{11}
\Delta_\pm r&=&\textstyle{1\over2}r\theta_\pm\\
\Delta_\pm f&=&\nu_\pm\\
\Delta_\pm\theta_\pm&=&-\nu_\pm\theta_\pm-\textstyle{1\over2}\theta_\pm^2
-8\pi T_{\pm\pm}\\
\Delta_\pm\theta_\mp&=&-\theta_+\theta_--e^{-f}r^{-2}+8\pi T_{+-}\\
\Delta_\pm\nu_\mp&=&-\textstyle{1\over2}\theta_+\theta_--e^{-f}r^{-2}
+8\pi(T_{+-}+\textstyle{1\over2}e^{-f}h^\sharp:T)
\end{eqnarray}
have the same form as in spherical symmetry.
These equations, the quasi-spherical equations, 
therefore determine a quasi-spherical background.
The second-level equations 
\begin{eqnarray}
\Delta_\pm k&=&r^{-2}\sigma_\pm\\
\Delta_\pm\sigma_\mp&=&
\textstyle{1\over2}(\theta_\pm\sigma_\mp-\theta_\mp\sigma_\pm)
+\sigma_+\cdot h^\sharp\cdot\sigma_-
+8\pi e^{-f}(\bot T-\textstyle{1\over2}(h^\sharp:T)h)
\label{1l}
\end{eqnarray}
constitute a wave equation for $k$, 
describing the gravitational-wave propagation,
as explained in detail in the following.
The quadratic shear term in the shear propagation equation (\ref{1l})
was omitted in the original reference\cite{qs}.
The third-level equations for $(\omega,s_\pm)$ 
need not be solved for the radiation problem.
This is because, fixing $u_+$ to be the outgoing direction, 
the Bondi news at null infinity $\Im^\pm$ is 
essentially $\sigma_\mp/r$\cite{mon}.
This determines the strain to be measured by a gravitational-wave detector,
as explained in the penultimate section.

In the second approximation, the truncated field equations also decouple, 
this time into only two levels,
with the last level for $(\omega,s_\pm)$ 
again being irrelevant to the radiation problem.
The remaining equations are 
\begin{eqnarray}
\label{21}
\Delta_\pm r&=&\textstyle{1\over2}r\theta_\pm\\
\Delta_\pm f&=&\nu_\pm\\
\label{2k}
\Delta_\pm k&=&r^{-2}\sigma_\pm\\
\Delta_\pm\theta_\pm&=&-\nu_\pm\theta_\pm-\textstyle{1\over2}\theta_\pm^2
-\textstyle{1\over4}|\sigma_\pm|^2-8\pi T_{\pm\pm}\\
\Delta_\pm\theta_\mp&=&-\theta_+\theta_--e^{-f}r^{-2}+8\pi T_{+-}\\
\Delta_\pm\nu_\mp&=&-\textstyle{1\over2}\theta_+\theta_--e^{-f}r^{-2}
+\textstyle{1\over4}\sigma_+:\sigma_-^\sharp
+8\pi(T_{+-}+\textstyle{1\over2}e^{-f}h^\sharp:T)\\
\Delta_\pm\sigma_\mp&=&
\textstyle{1\over2}(\theta_\pm\sigma_\mp-\theta_\mp\sigma_\pm)
+\sigma_+\cdot h^\sharp\cdot\sigma_-
+8\pi e^{-f}(\bot T-\textstyle{1\over2}(h^\sharp:T)h).
\label{2l}
\end{eqnarray}
The dual-null initial-data formulation is based on a spatial surface $S$
and the null hypersurfaces $\Sigma_\pm$ locally generated from $S$ 
in the $u_\pm$ directions.
The structure of the truncated field equations shows that one may specify 
$(\theta_\pm,r,f,k)$ on $S$,
$(\sigma_+,\nu_+)$ on $\Sigma_+$
and $(\sigma_-,\nu_-)$ on $\Sigma_-$.
In particular, the initial data is freely specifiable.
In summary, the vacuum system consists of 
nine first-order differential equations and their duals.

Mathematically, the difference between first and second approximations is that 
in the first approximation,
the equations for the quasi-spherical variables $(\theta_\pm,\nu_\pm,r,f)$
decouple from the equations for the wave variables $(\sigma_\pm,k)$.
Physically this describes gravitational-wave propagation 
on a quasi-spherical background.
The background is not fixed in advance
and need not be spherically symmetric,
so even the first approximation is widely applicable.
There is no such decoupling in the second approximation:
the gravitational-wave terms $(\sigma_\pm,k)$ 
now enter the equations for the quasi-spherical part of the geometry,
so there is no longer a background which is independent of the waves.
Physically this corresponds to including radiation reaction,
as clarified in the next section.

Nevertheless, 
both first and second approximations share the remarkable feature that,
to compute the observable waveforms,
no transverse $D$ derivatives need be considered.
The truncated equations form an effectively two-dimensional system,
to be integrated independently at each angle of the sphere.
Physically this means that the observed gravitational-wave signal 
depends only on the line of sight to the source, surely a plausible result.
Moreover, by virtue of the dual-null formulation,
the equations are already written in characteristic form,
the mathematically standard form for analysis of hyperbolic equations.
Numerical implementation is consequently straightforward 
and computationally inexpensive.
Numerical codes exist for both first and second approximations\cite{SH}.

\section{Gravitational radiation: local energy tensor}

The truncated equations in the second quasi-spherical approximation, 
(\ref{21})--(\ref{2l}),
differ from those of the first, (\ref{11})--(\ref{1l}), 
by terms quadratic in the shear tensors $\sigma_\pm$,
which appear additively with terms involving 
the energy tensor of the matter.
Specifically, the second approximation may be obtained from the first 
by replacing the matter energy tensor $T$ with $T+\Theta$, where
\begin{eqnarray}
\Theta_{\pm\pm}&=&\sigma_\pm:\sigma_\pm^\sharp/32\pi\\
\Theta_{+-}&=&0\\
\bot\Theta&=&e^f(\sigma_+:\sigma_-^\sharp)h/32\pi.
\end{eqnarray}
Thus $\Theta$ playes the role of an effective energy tensor 
for the gravitational radiation.
Written covariantly in terms of the transverse conformal metric $k$, 
this defines the {\em energy tensor of the gravitational radiation}:
\begin{equation}
\Theta_{\alpha\beta}={\langle\Delta_\alpha k,\Delta_\beta k\rangle
-{1\over2}g_{\alpha\beta}g^{\gamma\delta}
\langle\Delta_\gamma k,\Delta_\delta k\rangle\over{32\pi}}
\label{energy}
\end{equation}
where $\langle\alpha,\beta\rangle=k^{ab}k^{cd}\alpha_{ac}\beta_{bd}$
is the transverse conformal inner product
and $\Delta=dx^+\Delta_++dx^-\Delta_-$ is the 1-form evolution derivative.

More formally, 
in terms of the expansion parameters $\epsilon_0$ and $\epsilon_1$,
one may introduce a {\em truncated Einstein tensor} $C$ 
defined in terms of the full Einstein tensor $G$ by
\begin{eqnarray}
C_{\pm\pm}&=&\lim_{\epsilon_1\to0}\lim_{\epsilon_0\to0}G_{\pm\pm}
=-\Delta_\pm\theta_\pm-\nu_\pm\theta_\pm-\textstyle{1\over2}\theta_\pm^2\\
C_{+-}&=&\lim_{\epsilon_1\to0}\lim_{\epsilon_0\to0}G_{+-}
=\Delta_+\theta_-+\theta_+\theta_-+e^{-f}r^{-2}\\
h^\sharp:C&=&\lim_{\epsilon_1\to0}\lim_{\epsilon_0\to0}h^\sharp:G
=e^f(2\Delta_+\nu_--2\Delta_+\theta_--\theta_+\theta_-)\\
\tilde C&=&\lim_{\epsilon_0\to0}\tilde G
=e^f(\Delta_{(+}\sigma_{-)}-\sigma_+\cdot h^\sharp\cdot\sigma_-)
\end{eqnarray}
where $\tilde C=\bot C-{1\over2}(h^\sharp\cdot C)h$
and the $C_\pm$ components are irrelevant to the radiation problem.
Then 
\begin{equation}
C=8\pi T 
\end{equation}
are the truncated Einstein equations of the first approximation,
obtained from (\ref{11})--(\ref{1l}), whereas 
\begin{equation}
C=8\pi(T+\Theta)
\end{equation}
are the truncated Einstein equations of the second approximation,
obtained from (\ref{21})--(\ref{2l}).
This demonstrates that 
$\Theta$ is an energy tensor for the gravitational radiation,
in the sense that it is included as an effective matter energy tensor 
in the second approximation.
{\em Gravitational radiation reaction}, 
the back-reaction of the radiation on the space-time, 
has thereby been included.

It should be stressed that (i) mathematically, $\Theta$ is a genuine tensor,
but depends on the dual-null foliation, not just on the space-time;
(ii) the physical interpretation of $\Theta$ as energy requires 
the quasi-spherical approximation to be valid,
meaning that the transverse surfaces must indeed be roughly spherical.
This will not be made precise here,
as the range of validity of the approximation 
is not clear in advance and best explored in applications.
The intuitive meaning of roughly spherical should be clear by any standards.

In summary, the quasi-spherical approximation allows a local definition of
the energy-momentum-stress of gravitational radiation,
and therefore of the radiation itself:
{\em gravitational radiation} is present at a given point 
if and only if $\Theta$ is non-zero there.
With the orientation such that $u_+$ is the outgoing null direction, 
there is outgoing radiation if and only if 
$\Delta_-k$ (equivalently $\sigma_-$) is non-zero,
and ingoing radiation if and only if 
$\Delta_+k$ (equivalently $\sigma_+$) is non-zero.
The terminology gravitational radiation rather than wave 
is generally preferable,
since $\Delta_\pm k$ need not be oscillatory.
Instead, frequency spectra for ingoing and outgoing radiation 
may be defined by Fourier transformations 
to frequency $f_+$ and $f_-$ respectively:
\begin{equation}
k^\pm(f_\pm)=\int_\gamma e^{-2\pi if_\pm x^\pm}k(x^\pm)dx^\pm
\end{equation}
where $\gamma$ is a curve of constant $(x^\mp,\vartheta,\varphi)$.
If the Fourier transform is peaked in frequency space,
one may say that there is a {\em gravitational wave}.
In contrast, 
gravitational radiation is defined even when there is no typical frequency.
This clearly indicates that the approximation has a different physical basis 
to that of the Isaacson high-frequency approximation\cite{I,MTW}, 
which is usually quoted to make sense of linearized gravitational waves.

The non-zero components of $\Theta$ may be written as
\begin{eqnarray}
\Theta_{\pm\pm}&=&||\Delta_\pm k||^2/32\pi\\
\bot\Theta&=&e^f\langle\Delta_+k,\Delta_-k\rangle h/32\pi
\end{eqnarray}
where $||\alpha||^2=\langle\alpha,\alpha\rangle$ 
is the transverse conformal norm.
Then the $\Theta_{\pm\pm}$ components have a similar form
to those of the gravitational-wave energy tensor 
of the high-frequency approximation\cite{I,MTW},
with $k$ replacing the transverse traceless metric perturbation.
However, 
the high-frequency approximation requires averaging over several wavelengths 
and has no term proportional to $g$.
In this connection, it seems that the quasi-spherical situation allows 
a natural choice of transverse traceless gauge
and eliminates the need for averaging.
Earlier attempts to construct pseudotensors 
for gravitational waves by Einstein and others 
might be converted to genuine tensors
by similar gauge-fixing adapted to the transverse surfaces,
but currently such pseudotensors are generally not accepted.

Apart from the fact that a gravitational wave generally has two polarizations,
as encoded in the two independent components of $k$,
$\Theta$ is analogous to 
the energy tensor of Einstein-Rosen gravitational waves\cite{cyl},
which takes the massless Klein-Gordon form 
in terms of a gravitational potential generalizing the Newtonian potential.
In both cases, 
there is generally a transverse radiation pressure, given by $\bot\Theta$, 
produced by a combination of ingoing and outgoing gravitational radiation,
as well as the expected radial radiation pressure.
The fact that $\Theta_{+-}=0$ may be interpreted as meaning that 
gravitational radiation is purely radiative and workless,
as for the massless (but not massive) Klein-Gordon field.

As an energy tensor, $\Theta$ satisfies 
the strong, dominant, weak and null energy conditions\cite{HE}, as follows.
The dominant energy condition, 
requiring that an observer measures future-causal momentum,
may be stated as $T(\alpha,\beta)\ge0$ 
for two future-causal vectors $\alpha$, $\beta$.
Writing the vectors as
\begin{eqnarray}
\alpha&=&a^+l_++a^-l_-+(2e^{-f}a^+a^-)^{1/2}a\qquad\bot a=a\\
\beta&=&b^+l_++b^-l_-+(2e^{-f}b^+b^-)^{1/2}b\qquad\bot b=b
\end{eqnarray}
the future-causal conditions imply $a^\pm\ge0$, $b^\pm\ge0$, 
$|a|\le1$, $|b|\le1$ and therefore $0\le a\cdot b^\flat\le1$.
Then
\begin{equation}
32\pi\Theta(\alpha,\beta)
=a\cdot b^\flat\left|\left|(a^+b^+)^{1/2}\Delta_+k
+(a^-b^-)^{1/2}\Delta_-k\right|\right|^2
+(1-a\cdot b^\flat)
\left(a^+b^+||\Delta_+k||^2+a^-b^-||\Delta_-k||^2\right)\ge0.
\end{equation}
Thus $\Theta$ satisfies 
the dominant energy condition 
and therefore also the weak and null energy conditions.
The strong energy condition states that 
$\overline T(\alpha,\alpha)\ge0$ where $\overline T=T-{1\over2}(g^{-1}:T)g$.
Since
\begin{equation}
\overline\Theta_{\alpha\beta}={\langle\Delta_\alpha k,\Delta_\beta k\rangle
\over{32\pi}}
\end{equation}
this follows more simply as
\begin{equation}
32\pi\overline\Theta(\alpha,\beta)
=\left|\left|a^+\Delta_+k+a^-\Delta_-k\right|\right|^2\ge0.
\end{equation}
Thus by all standard measures, 
{\em gravitational radiation carries positive energy},
a physically expected property.

\section{Black holes: mass, surface gravity}

A general definition of dynamic black holes 
was proposed previously\cite{bhd,bhs}.
A transverse surface is said to be 
{\em trapped} if $\theta_+\theta_->0$,
{\em marginal} if $\theta_+=0$ or $\theta_-=0$
and {\em untrapped} if $\theta_+\theta_-<0$,
also previously called mean convex\cite{mon}.
Such definitions may also be applied independently at each spherical angle.
The surface is {\em future} trapped if $\theta_\pm<0$
and {\em past} trapped if $\theta_\pm>0$,
or {\em future} marginal if $\theta_-<0$ (for $\theta_+=0$)
and {\em past} marginal if $\theta_->0$.
An untrapped surface has a preferred spatial orientation\cite{mon}:
if $\theta_+>0$ and $\theta_-<0$,
then $\Delta_+$ is {\em outward} and $\Delta_-$ {\em inward}.
Similarly any spatial direction normal to the surface is outward or inward 
as its component along $\Delta_+$ is positive or negative respectively.
Future or past marginal surfaces similarly have a preferred spatial orientation.
A {\em trapping horizon} is a hypersurface foliated by marginal surfaces.
It is an {\em outer} trapping horizon if $L_-\theta_+<0$ (for $\theta_+=0$)
and an {\em inner} trapping horizon if $L_-\theta_+>0$,
where the dual-null foliation is adapted to the marginal surfaces.
A future (respectively past) outer trapping horizon is proposed 
as the local, dynamical definition of 
a non-degenerate black (respectively white) hole.
A degenerate black hole is one for which 
$\theta_+$ is decreasing in the $L_-$ direction, 
but $L_-\theta_+$ is not strictly negative.

These ideas of gravitational trapping may also be expressed 
in terms of the quasi-spherical radius $r$, 
as in spherical symmetry\cite{1st}, as follows.
A transverse surface is trapped, marginal or untrapped
as $\Delta^\sharp r$ is temporal, null or spatial respectively.
If $\Delta^\sharp r$ is future (respectively past) temporal or null, 
then the surface is future (respectively past) trapped or marginal.
On an untrapped or marginal surface,
an achronal (spatial or null) normal direction is outward or inward 
as $r$ is increasing or decreasing respectively.
A trapping horizon is outer or inner
as $\Delta^2r>0$ or $\Delta^2r<0$ respectively,
where $\Delta^2=\Delta^\sharp\cdot\Delta$.

In spherical symmetry, there is an active gravitational mass-energy 
with many desired properties\cite{1st,sph}.
A {\em quasi-spherical mass-energy} may be defined similarly as
\begin{equation}
m=\textstyle{1\over2}r(1-\Delta r\cdot\Delta^\sharp r).
\label{mass}
\end{equation}
This is the simplest generalization which has the property that 
a transverse surface is trapped, marginal or untrapped if and only if 
$r<2m$, $r=2m$ or $r>2m$ respectively.
Physically, $m$ includes the energy of the gravitational field,
as will be seen below in the first law.
Similarly, the spherically symmetric $m$ yields, in a post-Newtonian expansion,
the Newtonian mass, gravitational potential energy, 
kinetic energy and thermal energy of a thermodynamic material\cite{1st}.

A {\em canonical flow of time} is generated by the vector
\begin{equation}
\xi=({*}dr)^\sharp=({*}\Delta r)^\sharp
\end{equation}
where $*$ is the Hodge operator of the evolution space, 
${*}1=e^{-f}dx^+\wedge dx^-$.
One may define $\xi$ equivalently, up to sign, by
\begin{eqnarray}
\xi\cdot\Delta r&=&0\\
\xi\cdot\xi^\flat&=&-\Delta r\cdot\Delta^\sharp r\\
\bot\xi^\flat&=&0.
\end{eqnarray}
Then a sphere is trapped, marginal or untrapped 
as $\xi$ is spatial, null or temporal respectively.
In particular, trapping horizons may equivalently be defined as 
hypersurfaces where $\xi$ is null.
In these and other ways,
$\xi$ is analogous to the Killing vector of a stationary space-time
or the Kodama vector of a spherically symmetric space-time\cite{1st,sph}.

Denoting the Hodge operator of $g$ by $\star$, there is the truncation identity
\begin{equation}
\star1={*}1\wedge\bar{*}1={*}1\wedge\hat{*}r^2
\label{hodge}
\end{equation}
since the terms in the shifts $s_\pm$ are quadratic.
Then
\begin{equation}
\nabla\cdot\xi={\star}d{\star}\xi^\flat={\star}d(dr\wedge\hat{*}r^2)
=r^{-2}{*}d(r^2dr)=0.
\end{equation}
Thus $\xi$ is covariantly conserved.
This Noether current therefore admits a Noether charge 2-form $Q_\xi$
such that 
\begin{equation}
\xi^\flat={\star}dQ_\xi.
\end{equation}
Reversing the above calculation yields
\begin{equation}
Q_\xi=\hat{*}\textstyle{1\over3}r^3.
\end{equation}
Thus the integrated charge over a transverse surface is
\begin{equation}
\overline{V}=\oint Q_\xi=\textstyle{1\over3}\oint\hat{*}r^3
\end{equation}
which is the average radial volume.

A dynamic {\em surface gravity} $\kappa$ may be defined, 
generalizing that of the spherically symmetric case\cite{1st}, by
\begin{equation}
\top(\xi\cdot d\xi^\flat)=\kappa\Delta r
\end{equation}
where $\top$ indicates projection onto the evolution space.
That is, the 1-forms on each side of the equation are proportional
in the quasi-spherical truncation,
and $\kappa$ is defined as the proportionality constant.
Since $\xi^\flat=\pm\Delta r$ on a trapping horizon,
this reduces on such a horizon to
\begin{equation}
\top(\xi\cdot d\xi^\flat)=\pm\kappa\xi^\flat
\end{equation}
which is analogous to the usual definition of stationary surface gravity,
with $\xi$ replacing the stationary Killing vector.
A calculation shows that
\begin{equation}
\kappa=\textstyle{1\over2}\Delta^2r.
\label{sg}
\end{equation}
Therefore a trapping horizon is outer or inner
as $\kappa>0$ or $\kappa<0$ respectively.
Also, $\kappa$ vanishes where a black hole becomes degenerate,
a desired property of surface gravity.
Note that
$\kappa$ is defined everywhere in the space-time, not just on horizons.
Physically, $\kappa$ is a gravitational acceleration,
a relativistic version of the Newtonian gravitational acceleration 
or force per unit mass.

\section{Energy conservation: first law}

The truncated field equations (\ref{21})--(\ref{2l}) imply that
the mass-energy $m$ (\ref{mass}) is propagated as
\begin{equation}
\Delta_\pm m=4\pi r^2e^{f}\left(T_{+-}\Delta_\pm r
-(T_{\pm\pm}+\Theta_{\pm\pm})\Delta_\mp r\right)
\end{equation}
where the terms in $\Theta$ (\ref{energy}) appear 
in the second but not the first approximation.
This may be written more covariantly as the {\em unified first law} 
\begin{equation}
\Delta m=4\pi r^2(\psi+w\Delta r)
\label{law}
\end{equation}
in terms of the {\em work density} (an energy density)
\begin{equation}
w=e^fT_{+-}
\end{equation}
and the {\em energy flux} (a 1-form)
\begin{equation}
\psi=\psi[T]+\psi[\Theta]
\label{flux}
\end{equation}
which has been divided into contributions 
from the matter and the gravitational radiation:
\begin{eqnarray}
&&\psi[T]=\top(T\cdot\Delta^\sharp r)+w\Delta r\\
&&\psi[\Theta]=\top(\Theta\cdot\Delta^\sharp r).
\end{eqnarray}
The terms in the first law (\ref{law}) involving $\psi$ and $w$ 
may be interpreted as energy-supply and work terms respectively,
analogous to the heat supply and work 
in the classical first law of thermodynamics\cite{th}.
The unified first law was so called because it was shown in spherical symmetry 
that projecting it along the flow of a thermodynamic fluid 
yields a first law of relativistic thermodynamics,
while projecting it along a trapping horizon 
yields a first law of black-hole dynamics\cite{1st}.
It also includes the Bondi energy-loss equation at $\Im^+$,
with $m$ reducing to the Bondi mass,
$r^2\psi$ reducing to the Bondi flux and $w$ reducing to zero.
The quasi-spherical first law (\ref{law}) is unified in another sense:
it includes the energy flux of the gravitational radiation,
as also occurs for the first law in cylindrical symmetry\cite{cyl}.

The {\em energy-momentum density of the matter},
referred to the canonical flow of time, is the vector $j[T]$ given by 
\begin{equation}
j^\flat[T]=-\top(T\cdot\xi).
\end{equation}
In spherical symmetry, the analogous $j[T]$ is also conserved\cite{1st,sph}.
This also holds in the first but not second quasi-spherical approximation,
the physical reason being that gravitational radiation carries energy.
The {\em energy-momentum density of the gravitational radiation} 
is the vector $j[\Theta]$ given by
\begin{equation}
j^\flat[\Theta]=-\top(\Theta\cdot\xi).
\end{equation}
The energy fluxes are essentially duals of the energy-momentum densities:
\begin{eqnarray}
&&{*}j^\flat[T]=\psi[T]+w\Delta r\\
&&{*}j^\flat[\Theta]=\psi[\Theta].
\end{eqnarray}
It follows from the first law that
the combined energy-momentum density
\begin{equation}
j=j[T]+j[\Theta]
\end{equation}
is given by
\begin{equation}
j={({*}dm)^\sharp\over{4\pi r^2}}
={({*}\Delta m)^\sharp\over{4\pi r^2}}.
\end{equation}
Then, again using the truncation identity (\ref{hodge}), 
$j$ is covariantly conserved:
\begin{equation}
\nabla\cdot j={\star}d{\star}j^\flat
={{\star}d(dm\wedge\hat{*}1)\over{4\pi}}
={{*}ddm\over{4\pi r^2}}=0.
\label{cons}
\end{equation}
Physically this represents {\em conservation of energy}
for the gravitational radiation and matter combined.
The Noether charge 2-form $Q_j$ associated with the Noether current $j$ 
is given by
\begin{equation}
j^\flat={\star}dQ_j.
\end{equation}
Reversing the above calculation yields
\begin{equation}
Q_j={\hat{*}m\over{4\pi}}
\end{equation}
or essentially the mass-energy $m$.
The integrated charge over a transverse surface is 
\begin{equation}
\overline{m}=\oint Q_j={1\over{4\pi}}\oint\hat{*}m
\end{equation}
which is the average mass-energy.
Thus the first law (\ref{law}) 
is a first integral of the energy conservation equation (\ref{cons}).
A general method for defining mass-energy by Noether charges 
has been recently suggested\cite{noe}.
In summary, there is an energy conservation law 
for the combined energy-momentum of the matter and gravitational radiation.
Physically this shows how 
{\em energy may be locally transferred 
between matter and gravitational radiation},
the balance being accounted by the active gravitational mass-energy $m$.

\section{Black-hole dynamics: laws, inequalities}

A general local theory of black-hole dynamics 
was initiated a few years ago\cite{bhd} and recently reviewed\cite{mg9}.
The {\em second law of black-hole dynamics}\cite{bhd} states that 
the area element of a future outer trapping horizon is non-decreasing,
assuming the null energy condition.
Thus
\begin{equation}
r'\ge0
\label{l2}
\end{equation}
where the prime denotes 
the derivative along a vector generating the marginal surfaces, 
with orientation such that, when $r'=0$, it is future-null.
This is related to the {\em signature law}\cite{bhd},
that an outer trapping horizon is achronal (spatial or null),
being null if and only if $r'=0$.
The local second law implies an integral version: 
the area $A=\oint\hat{*}r^2$ of the marginal surfaces is also non-decreasing,
$A'\ge0$.
Since $r=2m$ on a trapping horizon, the second law also implies
\begin{equation}
m'\ge0
\end{equation}
so that the black-hole mass is also non-decreasing.
A similar {\em monotonicity} property of $m$ holds in an untrapped region,
as follows.
It is straightforward to show that 
$\psi[\Theta]$ is past (respectively future) causal 
in future (respectively past) trapped regions,
and outward achronal in untrapped regions.
The same holds for $\psi[T]$ assuming the null energy condition.
Also, the dominant energy condition implies $w\ge0$.
Therefore projecting the first law (\ref{law}) 
along an outward achronal direction, $m$ is non-decreasing in that direction.
This generalizes 
the monotonicity property of $m$ in spherical symmetry\cite{sph} 
and correspondingly yields two important inequalities.
Firstly, {\em positivity}:
in an untrapped region achronally outward from a regular centre,
\begin{equation}
m\ge0
\end{equation}
since $m$ vanishes at a regular centre.
Secondly, 
in an untrapped region achronally outward from a marginal surface with $r=r_0$,
\begin{equation}
m\ge r_0/2
\label{ineq}
\end{equation}
which is a local version of the Penrose inequality for black holes.
These inequalities extend to the asymptotic mass at $\Im^\pm$ 
or spatial infinity $i^0$ in an asymptotically flat space-time,
as in spherical symmetry\cite{sph}.

An explicit expression for the surface gravity (\ref{sg})
follows from the truncated equations 
of either first or second approximations, (\ref{21})--(\ref{2l}):
\begin{equation}
\kappa={m\over{r^2}}-4\pi rw.
\label{sg2}
\end{equation}
In vacuo, this has the same form as Newton's inverse-square law of gravitation,
combined with local equivalence of inertial and passive gravitational mass.
The dominant energy condition yields the inequality
\begin{equation}
m\ge r^2\kappa.
\end{equation}
Also, in an untrapped region or on a trapping horizon, $r\ge 2m$,
\begin{equation}
\kappa\le{1\over{2r}}.
\end{equation}
This means that, for a black hole of given size,
the surface gravity cannot exceed a given amount.
Both inequalities generalize those given in spherical symmetry\cite{1st}.
They may also be combined with the Penrose inequality (\ref{ineq}) 
and generalized to include charge\cite{in}.

Since $(m/r)'=0$ along a trapping horizon, (\ref{sg2}) implies
\begin{equation}
m'={\kappa\over{8\pi}}(4\pi r^2)'+w(4\pi r^2r').
\label{l1}
\end{equation}
This is the {\em first law of black-hole dynamics},
the projection of the unified first law (\ref{law}) along a trapping horizon.
It has the same form as that of spherical symmetry\cite{1st},
involving the radial area $4\pi r^2$ 
rather than the actual area $A$ of a marginal surface.
This suggests that dynamic black holes may in some sense 
admit a local Hawking temperature $\kappa/2\pi$ 
and a gravitational entropy $\pi r^2$\cite{HMA}.
It is noteworthy that, away from spherical symmetry, 
this entropy is generally not $A/4$.
However, this is consistent with the usual result for a Kerr black hole,
which has horizon area $A=4\pi r^2=4\pi(\tilde r^2+a^2)$,
where $\tilde r$ is the usual Kerr radial function\cite{SH}.

A {\em zeroth law of black-hole dynamics},
generalizing that of spherical symmetry\cite{mg9},
also follows from (\ref{sg2}):
if $r$ and $w$ are constant on a trapping horizon,
then so is the surface gravity $\kappa$.
Here constant $r$ is a local equilibrium condition for the horizon,
with constant $w$ being a similar condition for the matter.
It should be stressed that these three laws of black-hole dynamics 
all differ from the textbook laws of black-hole mechanics\cite{BCH,Wa},
for which the zeroth and first laws refer to stationary black holes,
specifically to a Killing horizon,
while the second law refers to a physically unlocatable concept, event horizon.

\section{Gravitational-wave dynamics: complex potential, wave equation}

The shear equations (\ref{2k}), (\ref{2l}), 
composed into a second-order equation for the transverse conformal metric $k$, 
become, taking the vacuum case,
\begin{equation}
\Delta_{(+}\left(r^2k^{-1}\cdot\Delta_{-)}k\right)=0.
\label{wave}
\end{equation}
This equation describes the propagation of the gravitational radiation,
or the {\em gravitational-wave dynamics}.
The original reference\cite{qs} omitted the $k^{-1}$ factor 
as a consequential error from (\ref{1l}).
To gain more insight into this wave equation,
it is useful to decompose the gravitational radiation 
into two independent polarizations.
A convenient choice of variables turns out to be two functions $(\phi,\chi)$
defined by
\begin{equation}
k=e^{-2\phi}\sec2\chi\,d\vartheta\otimes d\vartheta
+2\tan2\chi\sin\vartheta\,d\vartheta\otimes d\varphi
+e^{2\phi}\sec2\chi\sin^2\vartheta\,d\varphi\otimes d\varphi.
\end{equation}
Then $\phi$ and $\chi$ respectively encode 
the so-called plus and cross polarizations\cite{MTW} 
of the gravitational radiation,
referred to the quasi-spherical polar coordinates $(\vartheta,\varphi)$.
Straightforward calculations show that 
the gravitational-radiation energy tensor (\ref{energy}) takes the form
\begin{equation}
\Theta_{\alpha\beta}=
{\Delta_\alpha\phi\Delta_\beta\phi
+\Delta_\alpha\chi\Delta_\beta\chi
-\textstyle{1\over2}g_{\alpha\beta}g^{\gamma\delta}
(\Delta_\gamma\phi\Delta_\delta\phi
+\Delta_\gamma\chi\Delta_\delta\chi)
\over{4\pi\cos^22\chi}}.
\end{equation}
This is an energy tensor for two coupled scalar fields $(\phi,\chi)$,
in particular reducing to the Klein-Gordon form for purely plus polarization,
$\chi=0$.
These two potentials may be combined into 
a single complex {\em gravitational-radiation potential} 
\begin{equation}
\Phi=\phi+i\chi.
\label{pot}
\end{equation}
Then the energy tensor takes the form
\begin{equation}
\Theta_{\alpha\beta}=
{\Delta_\alpha\Phi\Delta_\beta\bar\Phi
-\textstyle{1\over2}g_{\alpha\beta}g^{\gamma\delta}
\Delta_\gamma\Phi\Delta_\delta\bar\Phi\over{4\pi\cosh^2(\Phi-\bar\Phi)}}
\end{equation}
where the bar denotes the complex conjugate.
The overall factor involving the cross polarization $\Phi-\bar\Phi$ indicates 
that this is not simply the the energy tensor of 
a complex Klein-Gordon field $\Phi$;
there is a non-trivial coupling.
Nevertheless, 
the gravitational radiation has been encoded in a complex potential,
acting like a complex scalar field.
This is analogous to the gravitational potential in cylindrical symmetry,
which also reduces to the Newtonian potential in the Newtonian limit\cite{cyl}.
The current case differs in that 
$\Phi$ is a potential for the gravitational radiation only,
not including the quasi-spherical part of the geometry,
for which one might instead define a gravitational potential $-m/r$.

The wave equation (\ref{wave}), written explicitly in terms of $(\phi,\chi)$ 
by straightforward calculations, becomes
\begin{eqnarray}
\Delta_+\Delta_-\phi
&+&r^{-1}(\Delta_+r\Delta_-\phi+\Delta_+\phi\Delta_-r)
+2\tan2\chi(\Delta_+\chi\Delta_-\phi+\Delta_+\phi\Delta_-\chi)=0\\
\Delta_+\Delta_-\chi
&+&r^{-1}(\Delta_+r\Delta_-\chi+\Delta_+\chi\Delta_-r)
+2\tan2\chi(\Delta_+\chi\Delta_-\chi-\Delta_+\phi\Delta_-\phi)=0.
\end{eqnarray}
Written covariantly in terms of $\Phi$ and the quasi-spherical wave operator
\begin{equation}
\Box=\Delta^2+2r^{-1}\Delta r\cdot\Delta^\sharp
\end{equation}
this becomes
\begin{equation}
\Box\Phi=2\tanh(\Phi-\bar\Phi)\Delta\Phi\cdot\Delta^\sharp\Phi.
\label{ernst}
\end{equation}
This complex wave equation encodes the gravitational-wave dynamics.
For purely plus polarization, $\Phi$ real,
the right-hand side vanishes and the equation is a linear wave equation, 
with the same form as the wave equation $\nabla^2\phi=0$ in spherical symmetry.
Otherwise, the cross polarization enters the equation 
and consequently affects the wave propagation.
This polarization-coupling effect is quadratic in $\Delta\Phi$ 
and so will be small for weak waves,
thereby agreeing with linearized gravitational-wave theory.
It should perhaps be stressed that,
while the gravitational-radiation potential $\Phi$ depends on 
all four space-time coordinates,
the wave equation (\ref{ernst}) is an effectively two-dimensional equation,
describing wave propagation independently at each angle of the sphere.
Similarly, while the wavefronts are roughly spherical,
the waves themselves need not be;
their amplitude is independent at each angle of the sphere.

Remarkably, the wave equation (\ref{ernst}) is a type of Ernst equation,
originally discovered in the quite different context of 
stationary axisymmetric space-times,
with $2\Phi$ being what Ernst denoted by $\mu$\cite{E}.
Ernst introduced other potentials more commonly denoted\cite{C,apw,G} by
\begin{equation}
Z=e^{2\Phi}
\end{equation}
and
\begin{equation}
E=\tanh\Phi
\end{equation}
in terms of which the wave equation (\ref{ernst}) becomes
\begin{equation}
(Z+\bar Z)\Box Z=2\Delta Z\cdot\Delta^\sharp Z
\end{equation}
or
\begin{equation}
(E\bar E-1)\Box E=2\bar E\Delta E\cdot\Delta^\sharp E.
\end{equation}
The Ernst equation has been extensively studied 
both in stationary axisymmetry\cite{C} and in plane symmetry\cite{G},
so existing methods can be used to analyse the gravitational-wave dynamics.
The current case is more complex in that 
$r$ is generally a function of all four space-time coordinates
and is not known independently of $\Phi$ in the second approximation.
However, in the first approximation, 
say on a Schwarzschild black-hole background,
the equation can be treated by existing analytical methods.
In particular, a direct comparison with linearized gravitational waves
should yield insight into the non-linear effects.
This at least demonstrates that, even in the very simplest case,
the approximation is not some disguised version of linearized theory.
In summary, the master gravitational-wave equation (\ref{ernst}) describes 
generally non-linear gravitational-wave propagation
in physical circumstances where the wavefronts are roughly spherical.

\section{Conformal fields: localized Bondi flux, strain}

In order to compare with observation,
one needs to examine the gravitational radiation at large distances 
in an almost flat background,
classically described by 
Bondi-Penrose asymptotic theory\cite{B,BBM,S1,S2,P1,P2,PR}.
This is straightforward in terms of the quasi-spherical conformal factor
\begin{equation}
\Omega=r^{-1}.
\end{equation}
Then $\Omega=0$ defines conformal boundaries of the space-time, 
including future null infinity $\Im^+$ 
and past null infinity $\Im^-$ 
for an asymptotically flat space-time.
Then the conformally rescaled expansions and shears 
\begin{eqnarray}
\vartheta_\pm&=&\Omega^{-1}\theta_\pm\\
\varsigma_\pm&=&\Omega\sigma_\pm
\end{eqnarray}
are finite and generally non-zero at $\Im^\mp$.
Rewriting the truncated field equations (\ref{21})--(\ref{2l})
of the second approximation yields, 
taking the vacuum case, 
\begin{eqnarray}
\Delta_\pm\Omega&=&-\textstyle{1\over2}\Omega^2\vartheta_\pm\\
\Delta_\pm f&=&\nu_\pm\\
\Delta_\pm k&=&\Omega\varsigma_\pm\\
\Delta_\pm\vartheta_\pm&=&-\nu_\pm\vartheta_\pm
-\textstyle{1\over4}\Omega||\varsigma_\pm||^2\\
\Delta_\pm\vartheta_\mp
&=&-\Omega(\textstyle{1\over2}\vartheta_+\vartheta_-+e^{-f})\\
\Delta_\pm\nu_\mp
&=&-\Omega^2(\textstyle{1\over2}\vartheta_+\vartheta_-+e^{-f}
-\textstyle{1\over4}\langle\varsigma_+,\varsigma_-\rangle)\\
\Delta_\pm\varsigma_\mp
&=&\Omega(\varsigma_+\cdot k^{-1}\cdot\varsigma_-
-\textstyle{1\over2}\vartheta_\mp\varsigma_\pm).
\end{eqnarray}
These conformal field equations are more practically suited 
to obtaining the gravitational waveforms at large distances, 
and are those which have been numerically implemented\cite{SH}.
Similarly, the energy flux $\psi$ (\ref{flux}) may be rescaled to yield 
a quantity which is generally non-zero at $\Im^\pm$, the {\em conformal flux}
\begin{equation}
\varphi=\Omega^{-2}\psi.
\end{equation}
For the gravitational-radiation energy flux, the explicit expressions
\begin{equation}
\varphi_\pm[\Theta]=-{e^f\vartheta_\mp||\varsigma_\pm||^2\over{64\pi}}
\end{equation}
have the same form as those for the Bondi flux at $\Im^\mp$\cite{mon}.
Thus the Bondi flux has been localized.
The unified first law (\ref{law}), projected along $\Im^+$,
thereby reduces to a Bondi energy-loss equation.
The entire preceding theory may be reformulated straightforwardly
in terms of such conformal fields.
The companion article\cite{gwe} illustrates this by using the conformal fields,
whereas this article generally uses the more familiar physical fields,
such as radius $r$.

The physically observable quantity to be measured 
by a gravitational-wave detector is the strain tensor $\epsilon$,
which determines the displacements 
\begin{equation}
{\delta\ell\over\ell}=\epsilon(e,e)
\end{equation}
where, in Newtonian theory, 
$e$ is a Cartesian basis vector in the direction of displacement.
In Einstein theory, the displacements for a transverse vector $e$ are given by
\begin{equation}
{h(e,e)+\delta h(e,e)\over{h(e,e)}}={(\ell+\delta\ell)^2\over{\ell^2}}.
\end{equation}
As a transverse surface approaches a metric sphere at $\Im^\pm$,
$\delta\Omega/\Omega\to0$ and $k(e,e)\to1$, so the formula yields
\begin{equation}
{\ell\over{\delta\ell}}\delta k(e,e)\to2.
\end{equation}
This leads to the identification of $\epsilon$ with $\delta k/2$,
or more precisely with
\begin{equation}
{1\over2}\int_\gamma\bot L_\alpha k\,d\tau
={1\over2}\int_\gamma\Omega(a^+\varsigma_++a^-\varsigma_-)d\tau
\end{equation}
where $\gamma$ is a worldline normal to the transverse surfaces
and $\alpha=\partial/\partial\tau=a^+l_++a^-l_-$ 
is a vector tangent to $\gamma$.
Since the strain itself vanishes at $\Im^\pm$,
it is more practical to introduce a quantity 
which is finite at future null infinity $\Im^+$,
the {\em conformal strain tensor}
\begin{equation}
\varepsilon={1\over2}\int_\gamma\varsigma_-\,dx^-
\end{equation}
where $\gamma$ is now a null curve of constant $(x^+,\vartheta,\varphi)$.
Then
\begin{equation}
\epsilon={\varepsilon\over{r}}
\end{equation}
is the strain tensor at a large distance $r$ from the source.
Thus the variables of the quasi-spherical approximation 
are directly related to the observable strain.
There is no need to match with another far-zone approximation,
such as a linearized approximation.
For instance, graphs of the strain waveforms, 
$\varepsilon$ against $x^-$ at a given angle $(\vartheta,\varphi)$,
are produced as output by the existing numerical codes\cite{SH}.

A familiar asymptotic relation between strain and energy density is recovered:
the energy density of outgoing gravitational radiation ($\varsigma_+=0$) 
near $\Im^+$ is, taking the normalization $a^\pm=1/\sqrt{2}$,
\begin{equation}
\Theta(\alpha,\alpha)
={||\Delta_-\epsilon||^2\over{16\pi}}
\end{equation}
where one should remember that 
the two independent components of $\epsilon$ each contribute twice in the norm.
Regarding conventions, an unfortunate relic of linearized theory 
is a definition of amplitude of a gravitational wave 
as that of the relevant component of 
the transverse traceless metric perturbation\cite{MTW},
which would correspond here to $k$.
As above, this differs by a factor of two from the strain amplitude,
which is a much more natural definition of gravitational-radiation amplitude,
being the observable quantity.
This is also simply related to 
the amplitude of the gravitational-radiation potential $\Phi$ by
\begin{equation}
\Theta(\alpha,\alpha)
={\Delta_-\Phi\Delta_-\bar\Phi\over{8\pi\cosh^2(\Phi-\bar\Phi)}}
\end{equation}
and is equal for purely plus polarization, $\Phi=\bar\Phi$,
as for Einstein-Rosen gravitational waves\cite{cyl}.
Another reason for the latter convention is that 
the energy density has the same form and numerical factor,
$(\hbox{amplitude}\times\hbox{frequency})^2/8\pi$,
as an electromagnetic wave in natural units.

\section{Conclusion}

This article has described an astrophysically realistic approximation scheme
in which both gravitational radiation and black holes are locally defined, 
along with their physical attributes,
with each dynamically influencing the other.
The approximation holds where the gravitational wavefronts 
and black hole (or other astrophysical object) are roughly spherical.
For black holes, this is an extension of 
the general theory of black-hole dynamics\cite{bhd,bhs,mg9} 
and its previous formulations in spherical\cite{1st} 
and cylindrical symmetry\cite{cyl}.
In particular, local definitions of gravitational mass-energy $m$ (\ref{mass})
and surface gravity $\kappa$ (\ref{sg}) have been given,
satisfying zeroth, first (\ref{l1}) and second (\ref{l2}) laws 
of black-hole dynamics.
For gravitational radiation, this is a new kind of approximation,
quite different from the commonly used asymptotic 
and high-frequency (or merely linearized) approximations.
Most crucially, 
gravitational radiation is {\em localized} in this approximation,
something which is usually argued to be impossible in general 
and in other existing approximations,
unless one counts symmetric space-times\cite{cyl}.
The general non-localizability of gravitational waves 
has long been an obstacle to understanding 
the physics of their most interesting, dynamic, strong-gravity sources.
Remarkably, in the quasi-spherical approximation, 
the gravitational-wave dynamics is described simply by 
an effectively two-dimensional wave equation (\ref{ernst}).
This analytically tractable Ernst equation can be used to study 
non-linear wave-propagation effects, for instance close to a black hole.

A local energy tensor $\Theta$ (\ref{energy}) for the gravitational radiation 
has been defined,
taking a scalar-field form 
in terms of a complex gravitational-radiation potential (\ref{pot}).
Including $\Theta$ like a matter energy tensor 
in the truncated Einstein equations 
thereby describes gravitational radiation reaction, 
the back-reaction of the radiation on the space-time,
e.g.\ on the black hole producing it.
A reliability test is provided by comparing this second approximation 
with the first approximation\cite{qs},
in which radiation reaction is not included 
and the radiation propagates on a background.
In the second approximation, 
the gravitational-radiation energy tensor $\Theta$ 
enters a covariant conservation law, 
conservation of energy (\ref{cons}).
Specifically, with respect to a canonical flow of time,
the combined energy-momentum of the matter and gravitational radiation 
is a Noether current,
with the Noether charge being the mass-energy $m$.
The energy conservation law can be expressed as a local first law ({\ref{law})
equating changes in $m$ to work and energy-supply terms, 
including the energy flux (\ref{flux}) of the gravitational radiation.
This unified first law includes both the first law of black-hole dynamics
and a Bondi energy-loss equation.

It may be useful to summarize here how 
the Einstein equation has been simplified and physically understood.
Of the ten components of the full Einstein equation,
four components of the truncated equations appear to involve angular momentum
but are irrelevant to the radiation problem.
Two of the remaining components constitute the wave equation (\ref{ernst}) 
describing the propagation of the gravitational radiation.
The remaining four components are the quasi-spherical equations,
of which one is an integrability condition for the others,
at least in vacuo and for some matter fields, such as a Klein-Gordon field.
Of these, 
one is a relativistic version of Newton's inverse-square law of gravitation,
(\ref{sg2}), and two are equivalent, up to an integration constant,
to the first law (\ref{law}).
Thus a comprehensive physical picture has emerged:
the Einstein equation has been essentially reduced to 
a quasi-Newtonian law of gravitation, 
energy conservation and a propagation equation for gravitational radiation.

As an example of current interest to many researchers, 
the approximation can be used to describe gravitational radiation from
a roughly spherical but dynamically evolving black hole 
formed by binary black-hole coalescence.
The distorted black hole emits gravitational radiation, 
absorbs backscattered radiation, 
thereby increases in area and traps some outgoing radiation, 
changes shape accordingly and consequently emits more radiation, and so on.
This ongoing dynamical process may now be described in a local way
by physically interpretable equations,
involving physical quantities such as black-hole mass and surface gravity 
and gravitational-radiation energy flux.
Apart from such practical applications to gravitational-wave astronomy, 
this provides a rich arena 
in which to advance physical understanding of 
both gravitational radiation and black holes,
and the dynamical interaction between them.

\bigskip\noindent
Acknowledgements.
Thanks to Hisa-aki Shinkai and Jong Hyuk Yoon for discussions.
Research partly supported by the APCTP visiting program
and a research grant of Konkuk University.

\end{document}